\documentclass[11pt]{article}
\usepackage{graphicx}
\usepackage{amsmath}
\usepackage{amssymb}
\usepackage{times}
\usepackage{subfigure}
\newtheorem{theorem}{Theorem}

\newtheorem{lemma}[theorem]{Lemma}
\newtheorem{definition}[theorem]{Definition}

\newcommand{\smartparagraph}[1]{\vspace{0.5ex}\noindent {\bf #1}\allowspace}
\def\allowspace{\hskip 0em plus 1em minus 0em\relax}

\def\thisPaperTitle{A Generalized Two-Phase Analysis of Knowledge Flows in Security Protocols 
}

\title{\thisPaperTitle}

\author{Marten van Dijk \and Emina Torlak  \and Blaise Gassend  \and  Srinivas Devadas \\ MIT Computer Science and Artificial Intelligence laboratory, Cambridge, USA\\ $\{$marten, emina, gassend, devadas$\}$@mit.edu}
\date{}

\begin{document}

\sloppy

\maketitle

\begin{abstract}
We introduce {\em knowledge flow analysis}, a simple and flexible formalism for checking cryptographic protocols. Knowledge flows provide a uniform language
for expressing the actions of principals, assumptions
about intruders, and the properties of cryptographic
primitives.  Our approach enables a generalized two-phase analysis:  we extend the two-phase theory by identifying the necessary and sufficient properties of a broad class of cryptographic primitives for which the theory holds.  We also contribute a library of standard primitives and show that they satisfy our criteria.

\end{abstract}

\noindent
{\bf keywords:} security protocols, intruder detection.

\section{Introduction}

One area of major successes for formal methods has been the verification of security protocols. A number of specialized tools have been developed in the last decade that have exposed subtle flaws in existing protocols (see, e.g. \cite{Lowe97, CJM00}).  Many of these tools \cite{ClarkeETAL98UsingStateSpaceExplorationNatural, CJM00, DurginETAL04MultisetRewritingSecurity} use a two-phase approach to efficiently identify intrusion scenarios.

This paper presents {\em knowledge flow analysis}, a lightweight and flexible formalism for checking cryptographic protocols.  Our approach is based on a simple mathematical foundation that provides an extensible framework for two-phase analysis.  In particular, we generalize the two-phase theory of Clarke et al \cite{ClarkeETAL98UsingStateSpaceExplorationNatural} by identifying the properties of cryptographic primitives for which the theory holds.  We demonstrate the generality of our criteria by using them to build a library of standard cryptographic primitives:  public and symmetric encryption/decryption, signing, pair/set construction, nonce generation, and hashing.  The sample library can be  extended to include blind signing, certification, and many other functions from the rich class of primitives that satisfy our criteria.  

Our approach gives a uniform formalism for
expressing the actions of principals, assumptions on
intruders, and properties of cryptographic primitives. The dynamic behavior of the protocol is described by an initial state of knowledge, and a collection of rules that dictate how knowledge may flow amongst principals.  Protocol rules are embedded into the initial state of knowledge as values that can be composed and decomposed by a special {\em rule primitive}, which satisfies the two-phase criteria.   

The knowledge flow approach grew out of an effort to check a new cryptographic scheme  \cite{Gassend03PhysicalRandomFunctions, GassendETAL02ControlledPhysicalRandomFunctions}. Knowledge flow analysis described here was the final result of a series of incremental attempts at formalizing and checking their assumptions using the Alloy language and tool \cite{Jackson02AlloyTOSEM, Jackson00AutomatingFOL}. This process drew out {\em the single source axiom} which, to the best of our knowledge, has not been described before:  the security of cryptographic functions depends on the assumption that their fixed points are hard to compute.   
   
The rest of the paper is structured as follows.  Section 2 explains the key intuitions underlying the approach, using the Needham-Schroeder protocol \cite{NS78}  as an example. Section 3 gives the  mathematical foundations of the approach.  Section 4 formulates and proves the two-phase theory in the knowledge flow context.  Section 5 provides a mathematical characterization of the primitives to which the two-phase theory applies.  Section 6 presents a sample subset of these primitives. The paper closes with a discussion of related work and concluding remarks.

\section{Knowledge Flow Basics}
The key idea behind knowledge flow analysis is the observation that, at the most basic level, the purpose of a security protocol is to distribute knowledge among its legitimate participants.  A protocol is flawed if it allows an intruder to learn a value that is intended to remain strictly within the legitimate principals' pool of knowledge.   To gain more intuition about knowledge flows in security applications, consider the original Needham-Schroeder protocol \cite{NS78}:
\begin{eqnarray*}
 1. &&  \mbox{Alice transmits }  {\cal E}_{{\cal PK}(B)}(I_A,N_A) \mbox{ to Bob} \\
 2. &&  \mbox{Bob transmits }  {\cal E}_{{\cal PK}(A)}(N_A,N_B) \mbox{ to Alice} \\
 3. &&  \mbox{Alice transmits }  {\cal E}_{{\cal PK}(B)}(N_B) \mbox{ to Bob}. 
\end{eqnarray*}

We have two principals, Alice and Bob, each of whom has an initial supply of knowledge.  Alice's initial knowledge, for example, consists of her own public/private key pair ${\cal PK}(A)$/${\cal SK}(A)$, identity $I_A$, nonce $N_A$, and Bob's public key ${\cal PK}(B)$ and identity $I_B$.  The purpose of the protocol is to distribute the nonces between Alice and Bob in such a way that the following conditions hold at the end:  (i) Alice and Bob both know $N_A$ and $N_B$, and (ii) no other principal knows both nonces.     

To initiate the protocol, Alice first expands her pool of knowledge to include ${\cal E}_{{\cal PK}(B)}(I_A, N_A)$, an encryption of her identity and nonce with Bob's public key.  She then sends the cipher to Bob who decrypts it using his private key, ${\cal SK}(B)$.  At the end of the first step of the protocol, Bob's knowledge has increased to include the values ${\cal E}_{{\cal PK}(B)}(I_A, N_A)$ and $N_A$.  Bob performs the second step of the protocol by adding ${\cal E}_{{\cal PK}(A)}(N_A, N_B)$ to his current knowledge and sending the cipher to Alice.  She uses her private key to decrypt Bob's message and extract $N_B$.  By using $N_B$ and $PK(B)$, Alice can set up an authenticated and private channel with Bob as is done during the final step of the protocol in which Alice creates ${\cal E}_{{\cal PK}(B)}(N_B)$ and forwards it to Bob.  Both Alice and Bob now know the two nonces and share all the knowledge  except their secret keys.

Following the flow of knowledge in the Needham-Schroeder protocol provides a crucial insight underlying our analysis method.  Namely, a principal can learn a value in one of three ways;  he can
\begin{list}{$\cdot$}{\setlength{\itemsep}{-3pt} }
\item \textit{draw} the value at the start,  
\item \textit{compute} it using his current knowledge, or 
\item \textit{learn} it by communication.  
\end{list}
Our analysis treats the latter two ways of obtaining knowledge as equivalent.  Specifically, we can think of Alice's computing ${\cal E}_{{\cal PK}(B)}(I_A, N_A)$ as her learning it from a principal called \textit{Encryptor} whose initial pool of values includes all possible ciphers:  Alice sends the tuple $({\cal PK}(B), (I_A, N_A))$ to \textit{Encryptor} who responds by sending back the encryption of $(I_A, N_A)$ with ${\cal PK}(B)$.  

Treating cryptographic primitives as principals allows us to consider the total pool of knowledge to be \textit{fixed}.  That is, the set of all values before and after the execution of a security protocol is the same; the only difference is the distribution of those values among the principals.  Since we assume that principals never forget values, the set of principals who know a value at the end of a protocol session subsumes the set of principals who drew the value at the beginning.      

The goal of analyzing knowledge flows in a protocol is to verify that particular values never leak out of  the honest participants' pool of knowledge.  In other words, \textit{we are interested in analyzing the flow of knowledge from an intruder's perspective}.  This observation allows us to make sound simplifying assumptions that drastically reduce the effort needed to formalize a protocol in terms of knowledge flows:  
\begin{list}{$\cdot$}{\setlength{\itemsep}{1pt} }
\item We need not encode the flows of knowledge among the honest principals, such as  the flow which allows Alice to learn ${\cal E}_{{\cal PK}(A)}(N_A, N_B)$ from $Encryptor$.  Rather, we may assume that each honest principal draws all values in the total knowledge pool and specify protocols solely in terms of the intruders' knowledge flows (sections \ref{kf-basic-defs} and \ref{initial-knowledge}).  
\item We may model all adversaries, including the untrusted public network, with a single opponent whom we call $Oscar$.  The soundness of this approach is formally proved in section \ref{adversary}.  Intuitively, the approach makes sense if we note that the potential adversaries will be most effective when they collaborate and share knowledge among themselves.  Hence, we can replace the (collaboration of) adversaries with a single principal who possesses all their knowledge, without excluding any intrusion scenarios.           
\end{list}

In our example, the flow of knowledge from the intruder's perspective starts with the protocol initialization message ${\cal E}_{{\cal PK}(B)}(I_A, N_A)$, since Oscar needs no prior knowledge to learn the first cipher that Alice sends to Bob.  In general, because Oscar includes the untrusted public network, he learns the first message of the protocol for free, regardless of who its intended recipient and sender are: 
\begin{equation}\forall_{p\in \{a,b\}, p'\in \{a,b\}\cup O} \ [\emptyset \rightarrow {\cal E}_{{\cal PK}(p')}({\cal I}(p),{\cal N}(\epsilon,{\cal I}(p)))].\label{ns1}
\end{equation}
The variables $a$ and $b$ denote the honest principals (Alice and Bob), and the set $O$ stands for Oscar.  The notation ${\cal N}(\epsilon,{\cal I}(p))$ represents the nonce that the nonce primitive generated for the principal identified by ${\cal I}(p)$ using the random value $\epsilon$ as the seed.  For example, Alice's identity is ${\cal I}(a)=I_A$ and Alice's nonce is
${\cal N}(\epsilon,{\cal I}(a))=N_A$. The empty set means that Oscar needs no prior knowledge to learn ${\cal E}_{{\cal PK}(p')}({\cal I}(p),{\cal N}(\epsilon,{\cal I}(p)))$.

Once his pool of knowledge includes ${\cal E}_{{\cal PK}(B)}(I_A, N_A)$, Oscar learns the corresponding response, ${\cal E}_{{\cal PK}(A)}(N_A, N_B)$.  More generally\footnote{We use the parameter $v$ in $c$ instead of ${\cal N}(\epsilon,{\cal I}(p))$ because $p'$, the recipient of $c$, cannot conclusively determine that $v$ is, in fact, the nonce ${\cal N}(\epsilon,{\cal I}(p))$.},
\begin{eqnarray}
\forall_{p'\in \{a,b\}, p\in \{a,b\}\cup O, v\in V} \left[\{c\} \rightarrow {\cal E}_{{\cal PK}(p)}(v,{\cal N}(c,{\cal I}(p')))\right]  \mbox{ where } c = {\cal E}_{{\cal PK}(p')}({\cal I}(p),v). \label{ns2} 
\end{eqnarray}
The variable $V$ denotes the set of all values, or the fixed pool of knowledge.  Note that our formalization constrains the seed of Bob's nonce to be Alice's initialization message.  This is needed to establish that Bob's nonce was generated in the context of the protocol session started by Alice with ${\cal E}_{{\cal PK}(B)}(I_A, N_A)$.  The resulting correspondence between the nonces prevents our analysis from sounding false alarms when Oscar legitimately obtains two nonces from Alice and Bob by running a valid protocol session with each.

Oscar learns the final message, ${\cal E}_{{\cal PK}(B)}(N_B)$, as a consequence of knowing ${\cal E}_{{\cal PK}(A)}(N_A, N_B)$.  Formally,
\begin{eqnarray}
 \forall_{p\in \{a,b\}, p'\in \{a,b\}\cup O, v\in V}  \left[\{{\cal E}_{{\cal PK}(p)}({\cal N}(\epsilon,{\cal I}(p)),v))\}  \rightarrow {\cal E}_{{\cal PK}(p')}(v) \right].\label{ns3}
\end{eqnarray}

\section{Knowledge Flow Analysis} 

Knowledge flow analysis is based on a simple mathematical foundation.  This section formalizes the ideas outlined in the discussion of knowledge flow basics.  We describe how \textit{communication rules} direct knowledge flows (\ref{kf-basic-defs}), show that our treatment of primitives ensures a fixed pool of values (\ref{initial-knowledge}), and formulate the analysis problem in terms of Oscar's knowledge flows (\ref{adversary}).  

\subsection{Communicating Knowledge}\label{kf-basic-defs}

We denote the sets of all {\em principals} and {\em values} by $P$ and $V$.  A subset of $P\times V$ is a {\em state of knowledge} drawn from $K=2^{P\times V}$, the set of all possible states of knowledge.  For a given state of knowledge $k\in K$, we say that ``$p$ knows $v$'' if $(p,v)\in k$.  
\begin{definition} A tuple $(R,k_0)$ is a {\em knowledge flow} for $(P,V)$ directed by the {\em communication rules} $R\subseteq P\times V\times P\times K$ and originating from the state $k_0\in K$.
\end{definition}
A communication rule describes the conditions under which one principal may gain knowledge from another.  For example, the rule $(e, {\cal E}_{{\cal PK}(p_b)}(v), p_a, \{(p_a, {\cal PK}(p_b)), (p_a, v)\})$ states that the encryptor $e$ will tell the cipher ${\cal E}_{{\cal PK}(b)}(v)$ to the principal $p_a$ if $p_a$ knows $p_b$'s public key and the plaintext $v$.  

Note that our definition of a communication rule limits the class of protocols expressible in the knowledge flow framework.  In particular, our rules cannot be used to specify conditions under which information is {\em withheld} from a principal, such as ``$a$ will {\em not} tell $v$ to $b$ if $b$ knows $x$''.  To the best of our knowledge, no protocol proposed for practical use requires this form of expressiveness.  

Given a set of communication rules $R$, we say that $k'\in K$ is reachable from $k\in K$ via $R$ if $k'$ is the result of applying all rules in $R$ to $k$ at most once; i.e. $k'=f_R(k)$ where 
\begin{definition} $f_R: K \longrightarrow K$ such that
$$f_R(k)=k\cup \left\{ (p_a,v) : \begin{array}{l}
(p_b,v) \in k, k_a \subseteq k, \mbox{ and } (p_b,v,p_a,k_a) \in R,\\
\mbox{for some } p_b\in P \mbox{ and } k_a\in K \end{array} \right\}.$$ \label{fR-def}
\end{definition}

A state of knowledge $k_n$ is reachable in the context of a knowledge flow $(R,k_0)$ if $k_n=f_R^n(k_0)$. The {\em maximal state of knowledge} $f^*_R(k_0)$ is the limit of $k_n=f_R^n(k_0)$ as $n\rightarrow \infty$.  A state of knowledge $f^*_{R_\kappa}(\kappa)$  is {\em valid} for a knowledge flow $(R,k_0)$ if $R_\kappa \subseteq R$ and $\kappa\subseteq k_0$.  Since $f_R(k_0)$ is monotonically increasing in $R$ and $k_0$, any valid state of knowledge is a subset of the maximal state of knowledge. Hence, the maximal state of knowledge is also the smallest fixed point of $f_R$ which subsumes $k_0$. It is evident from Definition \ref{fR-def} that self-rules such as $r=(p,v,p,k_p)\in R$ do not affect the flow of knowledge:  $f_R(k)=f_{R-\{r\}}(k)$. We therefore assume that $R$ does not contain any self-rules.

\subsection{Initial Knowledge}\label{initial-knowledge}

For each value $v$, $Source(v)=\{p: (p,v)\in k_0\}$ defines the set of principals who draw $v$.  In the knowledge flow framework, a principal $p$ outside of $Source(v)$ can learn $v$ only by communicating with principals who know $v$.  We therefore treat cryptographic primitives, and other computationally feasible algorithms, as principals. For example, suppose that, in practice, $p$ can compute $v$ by applying the algorithm ${\cal A}$ to inputs $i_1, i_2, \ldots, i_n$.  We model ${\cal A}$ by adding the principal $A$ to $P$, the tuple $(A,v)$ to $k_0$, and the rule $(A,v,p,\{(p,i_1),(p,i_2), \ldots (p, i_n)\})$ to $R$.

Our treatment of primitives ensures that $Knowledge(k_0)=\{v: (p,v)\in k_0 \mbox{ for some } p\in P\}$
consists of {\em all} learnable values. Hence, $V$ is the same in the initial and the maximal state of knowledge,
\begin{equation} Knowledge(k_0) = Knowledge(f^*_R(k_0)), \label{v0} \end{equation}
which implies that we can safely restrict our analysis to the subset of $R$ which only involves values in  $Knowledge(k_0)$. 

We further simplify our approach by constraining $k_0$, and therefore $R$, according to standard security assumptions.  Specifically, we assume {\em the single source axiom} for values that are fixed points of cryptographic functions.  For example, if the primitive $h$ models a hashing function ${\cal H}$, then we assume that $\{h\} = Source(x)$ for all $x$ such that $x={\cal H}(x)$.  We thus model the assumption that solving the equation $x={\cal H}(x)$ is computationally hard by stating that no principal other than $h$ can draw $x$:

\begin{definition}[Single Source Axiom] Set $F_p$ is {\em fixed} for a principal $p$ 
if for each  $(p,v,p_a,k_a)\in R$ with $v\in F_p$, there exists an $x\in F_p$ such that $(p_a,x)\in k_a$. Fixed sets $F_p$ for $p\in P$ satisfy the single-source axiom if, for all $p\in P$, $p_a\in P$,  and $v\in V$, 
$[v\in F_p \mbox{ and } (p_a,v)\in k_0] \Rightarrow [p=p_a]$.
\end{definition}

The consequence of the single-source axiom is that no principal outside of $\{p\}=Source(v)$ can ever learn $v\in F_p$:

\begin{lemma} \label{restr} If the fixed sets $F_p$ for $p\in P$ satisfy the single-source axiom, then,  for all $p\in P$, $p_a\in P$,  and $v\in V$,
$[v\in F_p \mbox{ and } (p_a,v)\in f^n_R(k_0)] \Rightarrow [p=p_a \mbox{ and } (p_a,v)\in k_0]$. 
\end{lemma}

\noindent
{\em Proof.} We use induction on $n$. The case $n=0$ is equivalent to the single-source axiom. Suppose that the lemma holds for $n$, our induction hypothesis.  Let $v\in F_p$ and 
 $(p_a,v)\in f^{n+1}_R(k_0)$. According to Definition \ref{fR-def}, (i) $(p_a,v)\in f^n_R(k_0)$ and the lemma follows from the induction hypothesis, or (ii) there exists a $p_b\in P$ and $k_a\in K$ such that $(p_b,v)\in f^n_R(k_0)$, $k_a\subseteq f^n_R(k_0)$, and $(p_b,v,p_a,k_a)\in R$. From the induction hypothesis we infer that $p=p_b$ and $(p_b,v)\in k_0$ since $v\in F_p$ and $(p_b,v)\in f^n_R(k_0)$. Hence, $(p,v,p_a,k_a)\in R$ and, since $v\in F_p$ and $F_p$ is a fixed set for $p$, there exists an $x\in F_p$ such that $(p_a,x)\in k_a\subseteq f^n_R(k_0)$, which proves $p=p_a$ by the induction hypothesis. Notice that $(p_a,v)\in k_0$ because $p_a=p=p_b$ and $(p_b,v)\in k_0$. The lemma follows by induction on $n$.

\hfill
$\Box$

Together with Equation (\ref{v0}), Lemma \ref{restr} implies that $f^*_{R_F}(k_0)=f^*_{R}(k_0)\subseteq k_0 \cup [P\times (Knowledge(k_0)-F)]$, where $F$ is the union of all $F_p$ and
$$R_F = \left\{ (p_b,x,p_a,k_a)\in R : \{(p_b,x)\}\cup k_a  \subseteq k_0 \cup [P\times (Knowledge(k_0) - F)] \right\}.$$
Hence, we need to analyze only the knowledge flows characterized by $R_F$.

\subsection{Adversaries' Knowledge}\label{adversary}

Let $O\subseteq P$ be a group of collaborating adversaries. We collapse $O$ into a single principal $o$ using the following merging function:
$$\begin{array}{rcl}
Merge(p) &=& \left\{\begin{array}{l}o \mbox{ if } p\in O, \\ p \mbox{ if } p\not\in O
\end{array}\right.\\
Merge(k) &=& \{(Merge(p),v) : (p,v) \in k \}\\
Merge(r) &=& (Merge(p_b),v,Merge(p_a),Merge(k_a)) \mbox{ where } r = (p_b,v,p_a,k_a) \in R
\end{array}$$
The merging of adversaries does not rule out any attacks because $Merge(f^*_R(k_0))\subseteq f^*_{Merge(R)}(Merge(k_0))$.  We subsequently assume that $Merge$ is implied and use $P$, $R$, and $k_0$ to refer to $Merge(P)$, $Merge(R)$, and $Merge(k_0)$.

Security properties of protocols are expressed as predicates on the values known to Oscar in the maximal state of knowledge.  We therefore focus our analysis of knowledge flows to finding all the values in the projection of $f^*_{R_F}(k_0)$ on Oscar.  Specifically, we introduce the projection function $g$ and show that its smallest fixed point is the image of Oscar under $f^*_{R_F}(k_0)$.
\begin{definition} \label{defarrow} Let $X \rightarrow x$ or, more explicitly, $X \rightarrow_p x$ denote the existence of a rule $(p,x,o,k_\sigma)\in R_F$ for some $p\in P-\{o\}$ and $k_\sigma\in K$ with $X=\{v :(o,v)\in k_\sigma\}$.  We define  $g: 2^V \longrightarrow 2^V$ as
$$g(X)=X\cup \left\{ x : X_\sigma \rightarrow x 
\mbox{ for some } X_\sigma\subseteq X \right\}.$$
 The set of values reachable from $X$ is given by $g^*(X)$, which is the limit of $g^n(X)$ as $n\rightarrow \infty$.\label{gR-def}
\end{definition}

Since $f_R(k_0)$ is monotonically increasing in $R$,
Oscar's pool of values under $f^*_R(k_0)$ is maximized if (a) 
Oscar tells everything he knows to the honest principals and (b) the honest principals tell each other all values learnable in polynomial time--which, in our framework, are the values in $Knowledge(k_0)-F$. Formally, Oscar's final knowledge is maximized when $[(P-\{o\})\times (Knowledge(k_0)-F)] \subseteq f^*_R(k_0)$.  This is equivalent to assuming that $[(P-\{o\})\times (Knowledge(k_0)-F)]\subseteq k_0$ because  $k\subseteq f^*_R(k_0)$ implies that $f^*_R(k_0)=f^*_R(k_0\cup k)$. 

\begin{theorem}[Knowledge Flow Analysis] \label{arrow}
Let $[(P-\{o\})\times (Knowledge(k_0)-F)]\subseteq k_0$ and let $k_n=f^n_R(k_0)$. Then 
the set $X_n=\{v:(o,v)\in k_n\}$ has the property that $X_n=g^n(X_0)$.
\end{theorem}

\noindent
{\em Proof.} 
We use induction on $n$.  For $n=0$, $X_n=X_0=g^n(X_0)$.  
Let $X_n=g^n(X_0)$, our induction hypothesis. We now need to prove that $X_{n+1}=g^{n+1}(X_0)$. By the definition of $X_{n+1}$,  $x\in X_{n+1} \Longleftrightarrow (o,x)\in k_{n+1}=f_R(k_n)$.  According to Definition \ref{fR-def}, $(o,x)\in f_R(k_n)$ if and only if (i) $(o,x)\in k_n$, which is  equivalent to $x\in X_n$, or (ii) there exists a $p\in P$ and $k_\sigma\in K$ such that $(p,x)\in k_n$, $k_\sigma\subseteq k_n$, and $(p,x,o,k_\sigma)\in R$. Since there are no self-rules $(o,v,o,k_\sigma)\in R$, we know that $p\in P-\{o\}$. Since $[(P-\{o\})\times (Knowledge(k_0)-F)]\subseteq k_0\subseteq k_n$, $(p,x)\in k_n$ if and only if $(p,x)\in k_0\cup [P\times (Knowledge(k_0)-F)]$ by Lemma \ref{restr}. This argument also proves that   the condition $k_\sigma\subseteq k_n$ is equivalent to 
$$X_\sigma=\{v: (o,v)\in k_\sigma\} \subseteq \{v: (o,v)\in k_n\} =X_n \mbox{ and } k_\sigma\subseteq k_0\cup [P\times (Knowledge(k_0)-F)].$$ 
Notice that $\{(p,x)\}\cup k_\sigma \subseteq k_0\cup [P\times (Knowledge(k_0)-F)]$ gives us $(p,x,o,k_\sigma)\in R_F$. 
Therefore, case ii) holds if and only if there exists a set $X_\sigma\subseteq X_n$  such that $X_\sigma\rightarrow x$. By Definition \ref{gR-def}, case (i) or case (ii) holds if and only if $x\in g(X_n)$.  Hence, $X_{n+1}=g(X_n)$ and the theorem follows by induction on $n$.

\hfill $\Box$

\section{Two-Phase Theory} \label{two}

The formalism developed in the previous sections enables a systematic and efficient analysis of Oscar's knowledge flows.  Specifically, Oscar's final pool of values can be computed in {\em two phases} by first applying all the {\em decomposing} rules in $R$ and then all the {\em composing} ones.  This is a consequence of the `two-phase theory' \cite{DurginETAL04MultisetRewritingSecurity, ClarkeETAL98UsingStateSpaceExplorationNatural}, which we now formulate and prove in the knowledge flow framework.  


Intuitively, a {\em composing} rule combines its inputs into an output value from which some or all of the inputs can be extracted using a corresponding {\em decomposing} rule, if one exists.  For example, the composing rule $r_{z}=\{x,y\}\rightarrow_p x+iy$, where  $x\neq 0$ and $y\neq 0$, combines the non-zero real numbers $x$ and $y$ into the complex number $x+iy$.  The corresponding decomposing rules, $r_{x}=\{x+iy\}\rightarrow_p x$ and $r_{y}=\{x+iy\}\rightarrow_p y$, reconstruct the inputs to $r_{z}$ from its output.  Formally, we define composing and decomposing rules as follows:       

\begin{definition} \label{defcomp}
 Let $p$ be a principal with  a partial ordering $\prec_p$ on the set of values $V$. We call $X \rightarrow_p x$
\begin{itemize}
\item composing, if $X\prec_p x$, that is, $v\prec_p x$ for all $v\in X$, and
\item decomposing, if there exists a value $v\in X$ with $x\prec_p v$ such that $x\in X'$ for all composing $X'\rightarrow_p v$. We say that $v \mbox{ controls}_p \ x$.
\end{itemize}
Principal $p$ is composing/decomposing if there exists a partial ordering $\prec_p$ such that for all $X\subseteq V$ and $x\in V$,
$X \rightarrow_p x$ is composing or decomposing.
\end{definition}



Our definition permits the images of composing rules of different principals to intersect.  In practice, however, such intersections are hard to compute; that is, equations like ${\cal H}(z)={\cal E}_k(x)$, where ${\cal H}$ is a hashing and ${\cal E}$ an encryption function, cannot be solved in polynomial time.  We model this by assuming that the images of different principals' composing rules are disjoint:     
\begin{definition}[Global Collision Free Axiom]\label{global-cfa} Orderings $\prec_p$ are globally collision free if the sets $\{v: \exists_x \ [x\prec_p v]\}$ have empty intersections for different $p$.
\end{definition}
By Definition \ref{defcomp}, the set $\{v: \exists_x \ [x\prec_p v]\}$ is the image of the composing rules of a composing/decomposing principal $p$.  Hence, the global collision free axiom gives us the required condition that $[X \rightarrow_p v  \mbox{ is composing and } X' \rightarrow_{p'} v \mbox{ is composing}] \Rightarrow [p=p']$ for all $p$ and $p'$ in $P$, $X$ and $X'$ in $V$, and $v\in V$.


The two phase theory (Theorem \ref{cur}) follows immediately from Definitions \ref{defcomp} and \ref{global-cfa}.  It states that applying a decomposing rule after a corresponding composing rule yields no new information.  We can therefore derive Oscar's maximal state of  knowledge in a minimal number of steps by applying all the decomposing rules before their composing counterparts.

\begin{theorem}[Two-Phase Theory] \label{cur} Suppose that the orderings of principals in $P-\{o\}$ are  globally collision free: If $X\rightarrow x$ is decomposing, $v \mbox{ controls } x$, $v\in X$, $X'\neq \emptyset$, and $X'\rightarrow v$ is composing, then $x\in X'$. 
\end{theorem}

\noindent
{\em Proof.}
Suppose that  $v \mbox{ controls}_{p} \  x$, $X' \neq \emptyset$, and $X'\rightarrow_{p'} v$ is composing. Since $v \mbox{ controls}_{p} \  x$, $x\prec_{p} v$ and since $X'\rightarrow_{p'} v$ is composing, there exists a value $x'$ such that $x'\prec_{p'} v$.  The orderings $\prec_p$ and $\prec_{p'}$  are globally collision free, therefore $p'=p$. By the definition of $v \mbox{ controls}_p\ x$, $x\in X'$ because $X'\rightarrow_{p} v$ is composing.

\hfill $\Box$

\section{Composing/Decomposing Principals} \label{ordprin}

The applicability of the two-phase theory is not restricted by its formulation in terms of composing/decomposing principals.  This section presents a general criterion for identifying composing/decomposing principals which we use in the next section to demonstrate that both standard cryptographic primitives and protocol rules are composing/decomposing in our framework. 

We represent composing and decomposing rules with {\em locally collision free sets}.  This representation ensures that each decomposing rule has a corresponding composing rule (\ref{s1}) and  that the composing rules are locally free of collisions (\ref{s2})---i.e., for all $p$ in $P$, all $X$ and $X'$ subsets of $V$, and all $v\in V$, $[X \rightarrow_p v  \mbox{ is composing and } X' \rightarrow_{p} v \mbox{ is composing}] \Rightarrow [X=X']$.  


\begin{definition}[Local Collision Free Axiom] \label{local} A set $S\subseteq V^m$ is locally collision free if there exist sets $C$ and $D$, which are subsets of $\{1,\ldots, m\}$, such that there exist subsets $W_i\subseteq \{1,\ldots, m\}$, $i\in C\cup D$, with the following properties:
\begin{equation}
  \mbox{for all } i\in D \mbox{ there exists a } h\in C  \mbox{ such that } h\in W_i \mbox{ and } i\in W_h \label{s1}
\end{equation}
and
\begin{eqnarray}
&&  \mbox{for all } i,t\in C \mbox{ and for all } (x_1,\ldots,x_m), (y_1,\ldots,y_m) \in S, \nonumber \\
&& \mbox{ if } x_{i}=y_{t} \mbox{ then } \{x_j:j\in W_i\}=\{y_j : j\in W_t\}. \label{s2}
\end{eqnarray}
The image $Im(S)$ of $S$ is defined as the set of values  $x_i$ for some $(x_1,\ldots,x_i, \ldots x_m)\in S$ such that  $i\in C$.
\end{definition}

An example of a locally collision free set is 
$$S=\{(s,{\cal G}(s),x,{\cal E}_{{\cal G}(s)}(x),{\cal S}_s(x)) : s,x\in V\}\subseteq V^5,$$
where ${\cal G}$, ${\cal E}$, and ${\cal S}$ are injective functions free of collisions; that is, ${\cal G}(v)\neq {\cal E}_{{\cal G}(s)}(x)$, ${\cal G}(v)\neq {\cal S}_s(x)$, and ${\cal E}_{{\cal G}(s)}(x)\neq {\cal S}_v(y)$ for all $s$, $v$, $x$, and $y$ in $V$.
Let $D=\{3\}$ indicate the position in the tuples in $S$ which correspond to $x$ and let $C=\{2,4,5\}$ indicate the positions which correspond to ${\cal G}(s)$, ${\cal E}_{{\cal G}(s)}(x)$, and ${\cal S}_s(x)$.  Let $W_3=\{1,4\}$, $W_2=\{1\}$, $W_4=\{2,3\}$, and $W_5=\{1,3\}$. Since ${\cal G}$, ${\cal E}$, and ${\cal S}$ are injective functions with disjoint images, $S$ satisfies (\ref{s2}).  Condition (\ref{s1}) is satisfied by taking $h = 4\in C$ and $i = 3\in D$. 

The following theorem shows how a local collision free set leads to a composing/decomposing principal. Its proof is in Appendix \ref{proof}.

\begin{theorem}[Composing/Decomposing] \label{theo3}  Let $p\in P-\{o\}$ be a principal such that
\begin{eqnarray}
&& (p, v, p_a, k_a)\in R \mbox{ implies } \nonumber \\
&& \hspace{1.5cm} \mbox{there exists an } i\in C\cup D \mbox{ and } (x_1,\ldots,x_m)\in S \mbox{ such that } \nonumber \\
&& \hspace{1.5cm} v=x_i,  k_a=\{(p_a,x_j):j\in W_i\}, \label{p1}
\end{eqnarray}
where $S$ is local collision free with respect to $C$, $D$, and $W_i$, $i\in C\cup D$.
Let $F_p$ be the maximal\footnote{There exists a unique maximal fixed set since the union of two fixed sets is again a fixed set.} fixed set with $F_p\subseteq Im(S)$. Then,  $p$ is composing/decomposing; composing rules correspond to $i\in C$ and decomposing rules correspond to $i\in D$. 
The image of the composing rules is $\{ v: \exists_{x\in V} \ [x\prec_p v]\}\subseteq Im(S)$.
\end{theorem}


Applying Theorem \ref{theo3} to our example, we define the encryptor/decryptor/signer $e$ by the decomposing rule 
$(e,x,p, \{(p,s),(p,{\cal E}_{{\cal G}(s)}(x))\})$  and the composing rules
$(e, {\cal G}(s), p, \{(p,s)\})$, 
$(e, {\cal E}_{{\cal G}(s)}(x), p, \{(p,{\cal G}(s)),(p,x)\})$ and
$(e,{\cal S}_s(x),p,\{(p,s),(p,x)\})$.  The decomposing rule corresponds to the position $3\in D$ and models decryption; the composing rules correspond to the positions $2\in C$, $4\in C$, and $5\in C$  and model public key generation, encryption, and signing.  The principal $e$ is therefore composing/decomposing, and the two-phase theory holds for $\cal E$, $\cal G$, and $\cal S$.

The composing/decomposing theorem is compatible with the knowledge flow theorem if the fixed set $F_p$ satisfies the single source axiom.  In Appendix \ref{proof}, we show that this is equivalent to assuming that it is hard to solve the equation $x=w(x)$ where $w(x)={\cal S}_{a}({\cal E}_{b}({\cal G}({\cal E}_{{\cal G}({\cal S}_{c}(x))}(d))))$ is some function composed of ${\cal G}$, ${\cal E}$, and ${\cal S}$.   


\section{Primitives} \label{prim}

We now present a sample library of composing/decomposing primitives, which is sufficient for modeling a wide range of security protocols.  The library includes the standard cryptographic primitives: encryption/decryption, signing, pair/set construction, nonce generation, and hashing.  It also provides a special {\em rule primitive} that allows protocol rules to be modeled in the composing/decomposing pattern.       

\subsection{Cryptographic Primitives}

\smartparagraph{Encryption/Decryption}
Public key encryption \cite{MOV97} consists of a (probabilistic)  encryption algorithm, a decryption algorithm, and a (probabilistic) key-generating algorithm.  Given some security parameter, the key-generating algorithm generates a public-secret key pair. We model $p$'s  private key ${\cal SK}(p)$ as belonging to $p$'s initial knowledge, $(p,{\cal SK}(p))\in k_0$. We model the public key as a one-way function of the secret key, i.e., ${\cal PK}(p)={\cal G}({\cal SK}(p))$. Hence, one can compute a corresponding public key from the given secret key but not vice versa. Letting $e$ denote the principal representing public key encryption, we can express key-generation as follows:  for all $p\in P$ and $s\in V$,
$(e, {\cal G}(s), p, \{(p,s)\})\in R$.
If we project this family of rules on Oscar, we obtain
\begin{equation}
 \forall_{s\in V} \ [\{s\}\rightarrow {\cal G}(s)].\label{e1}
\end{equation}

Given a plaintext $x$ and a public key ${\cal G}(s)$, the encryption algorithm computes the ciphertext\footnote{If the algorithm is probabilistic (for example in ElGamal encryption) then the ciphertext ${\cal E}_{{\cal G}(s)}(x;r)$ is also a function of some random value $r$ (uniformly) drawn by the algorithm.} ${\cal E}_{{\cal G}(s)}(x)$.   Thus, the parameterized rule for encryption is, for all $p\in P$ and $s,x\in V$,
$(e, {\cal E}_{{\cal G}(s)}(x), p, \{(p,{\cal G}(s)),(p,x)\})\in R$.
 Projecting the rule on Oscar yields
\begin{equation}
 \forall_{s,x\in V} \ [\{x,{\cal G}(s)\} \rightarrow {\cal E}_{{\cal G}(s)}(x)].\label{e2}
\end{equation}

Given a ciphertext ${\cal E}_{{\cal G}(s)}(x)$ and the secret key $s$, the decryption algorithm computes the plaintext $x$. Hence, for all $p\in P$ and $s,x\in V$,
$(e,x,p, \{(p,s),(p,{\cal E}_{{\cal G}(s)}(x))\})\in R$
and
\begin{equation}
 \forall_{s,x\in V} \ [\{s,{\cal E}_{{\cal G}(s)}(x)\} \rightarrow x].\label{e3}
\end{equation}


\smartparagraph{Signing}
We can model digital signatures \cite{MOV97} by extending $e$ with the rules of the form 
$(e,{\cal S}_s(x),p,\{(p,s),(p,x)\})\in R$,
for all $p\in P$ and $s,x\in V$, which translate into
\begin{equation} \forall_{v\in V} \ [\{x,s\}\rightarrow {\cal S}_s(x)]. \label{sign}
\end{equation}
  In practice, the principal who receives $x$ and its signature $S_s(x)$ can verify the signature by using the public key ${\cal G}(s)$. In our model, it is sufficient to note that knowledge of $y={\cal S}_s(x)$ already verifies that $y$ is a signature of $x$, signed by using the secret key $s$.  That is, the principal who obtained $y$ from $e$ knows both $s$ and $x$.

Symmetric key encryption is modeled by (\ref{e2}) and (\ref{e3}) where $G(s)$ is replaced by $s$ and where $s$ represents the symmetric key. We can extend this definition with (\ref{sign}) to include message authentication codes (MACs). 


\smartparagraph{Pairing/Set Construction}
Let $t$ be a principal such that a communication rule 
$(t,v,p_a,k_a)$ is in $R$ if and only if one of the following holds for $v\in V$, $p_a\in P$, and $k_a\in K$:
(i) $v=(x,y)$ and $k_a=\{(p_a,x),(p_a,y)\}$  for some $x,y \in V$, (ii) $k_a=\{(p_a,(v,y))\}$ for some $y \in V$, or (iii) $k_a=\{(p_a,(x,v))\}$ for some $x \in V$. 
Projected on Oscar, this set of rules becomes
\begin{eqnarray}
&& \forall_{x,y\in V} \ [\{x,y\} \rightarrow (x,y)],\label{d1} \\
&& \forall_{x,y\in V} \ [\{(x,y)\} \rightarrow x],\label{d2} \\
&& \forall_{x,y\in V} \ [\{(x,y)\} \rightarrow y]. \label{d3}
\end{eqnarray}
Replacing $(x,y)$ by $\{x,y\}$ in (\ref{d1}-\ref{d3}) turns $t$ into a primitive that generates sets of cardinality 2.

\smartparagraph{Nonce Generation}
Let ${\cal I}(p)\in V$ represent the public identity of $p$ (the identity function ${\cal I}$ embeds $P$ in $V$).  We model nonce generation with the nonce primitive $n$ and the parameterized rule 
$\forall_{p\in P, v\in V}(n, {\cal N}(v,{\cal I}(p)), p, \{(p,v)\})\in R$,
which translates into
\begin{equation} \forall_{p\in O, v\in V} \ [\{v\} \rightarrow {\cal N}(v,{\cal I}(p))]. \label{st1}
\end{equation}
The dependence of $n$'s output on ${\cal I}(p)$ ensures that $p$ cannot learn other principals' nonces from $n$. The parameter  $v$ represents the seed from which a pseudo random nonce is generated. If a  protocol stipulates that a principal $p$ needs to generate a new nonce in response to a received message $m$, then $v$ is taken to be equal to $m$. For the first nonce of a protocol, we take $v$ to be the empty string $\epsilon$.  

\smartparagraph{Hashing}
We define the primitive $h$ for calculating hashes ${\cal H}(x)$ with the family of rules $\forall_{p\in P, x\in V} \ (h, {\cal H}(x), p, \{(p,x)\})\in R$
and
\begin{equation}
\forall_{x\in V} \ [\{x\}\rightarrow {\cal H}(x)].\label{cp1}
\end{equation}

\subsection{The Rule Primitive} \label{prot}

Protocol rules do not compute new values; rather, they model the transmission of values computed by the primitives.  We can therefore embed protocol rules into the initial state of knowledge as follows. 

With each (parameterized) protocol rule $x\rightarrow y$, we associate the value $|x\rightarrow y|\in V$. This value is composed/decomposed by the {\em rule primitive} $r$ via the parameterized rules 
$$ \forall_{p\in P, x,y\in V} \ (r,|x\rightarrow y|,p, \{(p,x),(p,y)\})\in R \mbox{ and }$$
$$ \forall_{p\in P, x,y\in V} \ (r,y,p,\{(p,|x\rightarrow y|),(p,x)\})\in R,$$
whose projected forms are 
\begin{equation} \forall_{x,y\in V} \ [\{x,y\} \rightarrow |x\rightarrow y|]  \mbox{ and }
\label{eqr1}
\end{equation}
\begin{equation} \forall_{x,y\in V} \ [\{x,|x\rightarrow y|\} \rightarrow y].
\label{eqr2}
\end{equation}


We represent each protocol rule $x\rightarrow y$ with the initial knowledge $(o,|x\rightarrow y|)$. Oscar can now use (\ref{eqr2}) to learn $y$ from $|x\rightarrow y|$ if he knows $x$.  For example, the following addition to $k_0$, together with (\ref{eqr2}), simulates the rule (\ref{ns3}) of the Needham-Schroeder protocol:
\begin{eqnarray}
 && \forall_{p\in \{a,b\}, p'\in \{a,b\}\cup O, v\in V}  \left[(o,|\{{\cal E}_{{\cal PK}(p)}({\cal N}(\epsilon,{\cal I}(p)),v))\}  \rightarrow {\cal E}_{{\cal PK}(p')}(v)|)\in k_0 \right].
\nonumber
\end{eqnarray}

\subsection{Summary} \label{coll}

The rules (\ref{e1})-(\ref{eqr2}) define a library of primitives---$e$, $t$, $n$, $h$ and $r$---that are composing/decomposing according to Theorem \ref{theo3}.  The following assumptions are implicit in their definitions:  

\begin{list}{$\cdot$}{\setlength{\itemsep}{-3pt} }
\item[\textbf{The Collision Free Axioms}] It is hard to compute collisions of the composition rules  (\ref{e1}), (\ref{e2}), (\ref{sign}), (\ref{d1}), (\ref{st1}), (\ref{cp1}) and (\ref{eqr1}). Therefore, we model these rules as injective functions that are mutually free of collisions; that is, they satisfy the local and global collision free axioms.

\item[\textbf{The Single Source Axiom}] It is hard to compute fixed points of functional compositions of the  rules (\ref{e1}), (\ref{e2}), (\ref{sign}), (\ref{d1}), (\ref{st1}), (\ref{cp1}) and (\ref{eqr1}).  The principals $e$, $t$, $n$, $h$ and $r$ hence satisfy the single-source axiom.

\item[\textbf{Cryptographic Primitive Properties}] It is hard to compute the inverses that are not encoded by the decomposition rules (\ref{e3}), (\ref{d2}), and (\ref{d3}). The rules (\ref{e1}-\ref{d3}) represent Oscar's computational means in The Dolev-Yao intruder model \cite{DY83}. This assumes perfect cryptography: the set of values is supposed to be a free algebra. 
\end{list}

Collision freeness and perfect cryptography are routinely assumed when reasoning about security protocols.  To the best of our knowledge, however, the necessity of assuming the single source axiom has not been recognized before.  We discovered it using the Alloy Analyzer \cite{Jackson00AutomatingFOL}, a general purpose model finder, to check a security theorem about knowledge flows in the CPUFs renewal protocol \cite{Gassend03PhysicalRandomFunctions, GassendETAL02ControlledPhysicalRandomFunctions}:  in the absence of the axiom, the Analyzer generates a false counterexample to the theorem based on a fixed value that satisfies the equation $x={\cal E}_s(x)$.

\section{Related Work}

The first formalisms
designed for reasoning about cryptographic protocols are belief logics such as
BAN logic \cite{BurrowsETAL90LogicAuthentication},
used by the Convince tool 
\cite{LichotaETAL96VerifyingCryptographicProtocolsElectronicCommerce} 
with the HOL theorem prover \cite{GordonMelham93HOL}, and its generalizations (GNY \cite{GNY90}, AT \cite{AT91}, and  SVO logic \cite{SO94}
which the C3PO tool \cite{Dekker00C3PO} employs with the Isabelle theorem prover
\cite{NipkowETAL02IsabelleHOLTutorial}).
Belief logics are difficult to use since the logical form of a protocol does not correspond to the protocol itself in an obvious way.  Almost indistinguishable formulations of the same problem lead to different results. It is also hard to know if a formulation is over constrained or if any important assumptions are missing. BAN logic and its derivatives cannot deal with security flaws resulting from interleaving of protocol steps \cite{BM93} and cannot express any properties of protocols other than authentication \cite{MB93}.
To overcome these limitations, the knowledge flow formalism has, like other approaches \cite{Lowe97,MMS97,CJM00,SongETAL01AthenaNovelApproachEfficientAutomatic,Meadows94NRLProtocolAnalyzer}, a concrete operational model of protocol execution.  Our model also includes a description of how the honest participants in the protocol behave  and a description of how an adversary can interfere with the execution of the protocol. 

Specialized model checkers such as Casper \cite{Lowe97}, Mur$\phi$ \cite{MMS97},  Brutus \cite{CJM00}, TAPS \cite{Cohen04TAPS}, and ProVerif \cite{AB} have been successfully used to analyze security protocols.  These tools are based on state space exploration which leads to an exponential complexity. Athena \cite{SongETAL01AthenaNovelApproachEfficientAutomatic} is based on a modification of the strand space model \cite{FabregaETAL98StrandSpaces}. Even though it reduces the state space explosion problem, it remains exponential.  Multiset rewriting \cite{DurginETAL04MultisetRewritingSecurity} in combination with tree automata is used in Timbuk \cite{timbuk}. The relation between multiset rewriting and strand spaces is analyzed in \cite{MR-SP}. The relation between multiset rewriting and process algebras \cite{pi,spi} is analyzed in \cite{MR-PA}. 

Proof building tools such as NRL, based on Prolog \cite{Meadows94NRLProtocolAnalyzer}, have also been helpful for analyzing security protocols. However, they are not fully automatic and often require extensive user intervention. Model checkers lead to completely automated tools which generate counterexamples if a protocol is flawed. For theorem-proving-based approaches, counterexamples are hard to produce.

For completeness, we note that if the initial knowledge of the intruder consists of a finite number of explicit (non-parameterized, non-symbolic) values, then a polynomial time intruder detection algorithm can be shown to exist using a generalization of the proof normalization arguments
\cite{McAllester93AutomaticRecognitionTractabilityInferenceRelations,
BasinGanzinger01AutomatedComplexityAnalysisBasedOrderedResolution,
GivanMcallester02PolynomialInference}, 
which were employed in \cite{BodeiETAL02FlowDolevYao,
NielsonETAL01CryptographicAnalysisCubicTime} 
and have been implemented in the framework 
\cite{NielsonETAL04SuccinctSolverSuite} (our two phase theorem can also be used to derive a polynomial time algorithm). However, in practice,  the initial knowledge of an intruder is unbounded and represented by a finite number of parameterized sets, each having an infinite number of elements. 

\section{Concluding Remarks}

We introduced knowledge flow analysis, a new framework for reasoning about knowledge in cryptographic protocols. The key advantage of the knowledge flow approach over other formalisms is its simplicity and flexibility. It is simple in the sense that the underlying mathematics is straightforward and elementary; it does not require any specialized background (in logic). It is flexible in the sense that the same library of cryptographic primitives can be used to model different protocols and that the security of a complex scheme involving multiple protocols can be verified. Knowledge flow analysis allows modeling of confidentiality and authenticity via a wide range of primitives such as pairing, union, hashing, symmetric key encryption, public key encryption, MACs and digital signatures.

Our formalism derives its simplicity from being just sufficiently expressive to enable modeling of practical cryptographic protocols.  In particular, existentials \cite{DurginETAL04MultisetRewritingSecurity} cannot be encoded as knowledge flows; existentials are implicitly modeled in Oscar's initial knowledge.  NP-hardness proofs which use (existential) Horn clause reduction \cite{DurginETAL04MultisetRewritingSecurity} or SAT3 reduction  \cite{RT2001} are not applicable to knowledge flow analysis.

Our formalism leads to a rigorous mathematical treatment and generalization of the two-phase theory \cite{DurginETAL04MultisetRewritingSecurity, ClarkeETAL98UsingStateSpaceExplorationNatural} which is used to efficiently verify protocols. Our treatment reveals the necessary and sufficient  collision free and single source axioms; it is hard to compute collisions and fixed points of compositions of  cryptographic primitives. To the best of our knowledge the necessity of assuming the single source axiom has not been recognized before. 

\appendix
\section{Fixed Sets and Orderings} \label{proof}

To prove Theorem \ref{theo3}, we define a sequence of subsets which we use to define a partial ordering and  to characterize a fixed set.

\begin{definition}
Let $S$ be locally collision free with respect to $C$, $D$, and $W_i$, $i\in C\cup D$. We  define $S_n\subseteq V$ recursively by $S_{-1}=\emptyset$,
$$S_0=V- Im(S)=V-\{x_i :  i\in C \mbox{ and there exists a tuple } (x_1,\ldots,x_m)\in S \},$$
and, for $n\geq 0$,
$$S_{n+1}=S_n\cup \left\{ x_i : \begin{array}{l} i\in C \mbox{ and there exists a } (x_1,\ldots,x_m)\in S \\ \mbox{such that } x_j\in S_n \mbox{ for } j\in W_i \end{array} \right\}.$$
We define
$S_\infty = \{ v\in V : v\in S_n \mbox{ for some } n\geq 0 \}$.
\end{definition}

We first show in Lemma \ref{lemfixed} that $V- S_\infty$ is a fixed set. We start with a result which we use throughout the whole proof.

\begin{lemma} \label{lemcn} Let $i\in C$, $(x_1,\ldots, x_m)\in S$, and $x_i\in S_\infty$. Then (i) $x_i \in S_{n+1}- S_n$ for some $n\geq 0$ and (ii)  $x_j\in S_n$ for all $j\in W_i$ and there exists a $h\in W_i$ such that $x_h\in S_n- S_{n-1}$.
\end{lemma}

\noindent
{\em Proof.} (i) Since $i\in C$, $x_i\not\in S_0$ which proves $n\geq 0$. (ii) Let $k\geq 0$ be the smallest index for which  $x_j\in S_k$ for all $j\in W_i$. Then, there exists an index $h\in W_i$ such that $x_h\in S_k- S_{k-1}$. Notice that $x_i\in S_{k+1}$ by the definition of $S_{k+1}$. Therefore, if $k<n$  then $x_i\in S_{k+1}\subseteq S_n$, contradicting $x_i\in  S_{n+1}- S_n$. This proves $n\leq k$. 

From the definition of $x_i\in S_{n+1}- S_n$ we infer that there exists a $t\in C$, $(y_1,\ldots,y_m)\in S$ such that $x_i=y_t$ and $y_j\in S_n$ for $j\in W_t$. Since $i\in C$, $t\in C$, and $x_i=y_t$, (\ref{s2}) yields $x_h\in \{x_j:j\in W_i\}=\{y_j: j\in W_t\}\subseteq S_n$. From $x_h\in S_k- S_{k-1}$, we infer $n\geq k$. We conclude $n=k$ which proves the lemma.

\hfill $\Box$

\begin{lemma} \label{lemfixed} $V- S_\infty$ is a fixed set for $p$.
\end{lemma}

\noindent
{\em Proof.} Let $(p,v,p_a,k_a)\in R$ with $v\in V- S_\infty$. From (\ref{p1}) we infer that there exists an $i\in C\cup D$ and $(x_1,\ldots,x_m)\in S$ such that $v=x_i$ and $k_a=\{(p_a,x_j):j\in W_i\}$. If $i\in C$ and $x_j\in S_\infty$ for all $j\in W_i$, then, by the definition of $S_\infty$, $x_i\in S_\infty$, which contradicts $x_i=v\in V- S_\infty$. Hence, if $i\in C$ then  $x_j\in V- S_\infty$ for some $j\in W_i$. 

If $i\in D$, then
(\ref{s1}) shows the existence of  a $h\in C$ with $i\in W_h$ and $h\in W_i$.  From $i\in W_h$ and $h\in C$ we infer that if $x_h\in S_\infty$ then, by Lemma \ref{lemcn} (ii), $x_i\in S_\infty$, which contradicts $x_i=v\in V- S_\infty$. Hence,  $x_j\in V- S_\infty$ for some $j\in W_i$.
 
\hfill $\Box$

Let $F_p$ be the maximal fixed set such that $F_p\subseteq V- S_0=Im(S)$. Then Lemma \ref{lemfixed} proves that $V- S_\infty \subseteq F_p$, hence, $V- F_p \subseteq S_\infty$.
Notice that $X\rightarrow_p x$ (defined by using $R_F$) implies that there exists a rule $(p,x,o,k_\sigma)\in R$ such that $X=\{v:(o,v)\in k_\sigma\}$ and $X\cap F_p=\emptyset$. Since $F_p$ is a fixed set, $x\in F_p$ contradicts $X\cap F_p=\emptyset$. Thus, for $X\rightarrow_p x$, both $x\not\in F_p$ and  $X\cap F_p=\emptyset$, that is $x\in S_\infty$ and $X\subseteq S_\infty$. 

Sets $S_n$ lead to the partial ordering
\begin{eqnarray*}
 [v\prec_p w] \ &\equiv& 
\ [v\in S_a- S_{a-1} \mbox{ and } w\in S_b- S_{b-1} \mbox{ for some } 0\leq a<b].
\end{eqnarray*}
Notice that  $\{ v: \exists_{x\in V} \ [x\prec_p v]\}\subseteq V- S_0=Im(S)$.
Theorem \ref{theo3} follows from Lemmas \ref{lemcon}-\ref{lemdes}, which prove that $p$ is composing/decomposing. 

\begin{lemma} \label{lemcon} Let $(x_1,\ldots,x_m)\in S$ with $\{x_j:j\in W_i\}\rightarrow_p x_i$. Then, $\{x_j:j\in W_i\}\rightarrow_p x_i$ is composing if and only if $i\in C$.
\end{lemma}

 \noindent
{\em Proof.} Suppose that $i\in C$. Since $\{x_j:j\in W_i\}\rightarrow_p x_i$, $x_i\in S_\infty$.  By Lemma \ref{lemcn} (i), $x_i\in S_{n+1}- S_n$ for some $n\geq 0$, by Lemma \ref{lemcn} (ii),  $\{x_j:j\in W_i\} \prec_p x_i$, which proves that $\{x_j:j\in W_i\}\rightarrow_p x_i$ is composing.

Suppose that $\{x_j:j\in W_i\}\rightarrow_p x_i$ is composing, that is, $\{x_j:j\in W_i\}\prec_p x_i$.
If $i\in D$, then by (\ref{s1}) there exists a $h\in C$ with $h\in W_i$ and $i\in W_h$. 
Since $\{x_j:j\in W_i\}\rightarrow_p x_i$, $x_h\in S_\infty$ and by Lemma \ref{lemcn} (i),   $x_h\in S_{n+1}- S_n$ for some $n\geq 0$.  By Lemma \ref{lemcn} (ii), $x_i\in \{x_j:j\in W_h\}\subseteq S_n$, hence, $x_i \prec_p x_h$. This contradicts $\{x_j:j\in W_i\}\prec_p x_i$ and we conclude that  $i\not\in D$, that is, $i\in C$.

\hfill $\Box$

\begin{lemma} \label{lemdes} Let $(x_1,\ldots,x_m)\in S$ with $\{x_j:j\in W_i\}\rightarrow_p x_i$. Then, $\{x_j:j\in W_i\}\rightarrow_p x_i$ is decomposing if and only if $i\in D$.
\end{lemma}

 \noindent
{\em Proof.} We prove $\{x_j:j\in W_i\}\rightarrow_p x_i$ is decomposing for $i\in D$. Then, the lemma follows from Lemma \ref{lemcon} since $i\not\in D \Longleftrightarrow i\in C$ lead to composing rules.
 By (\ref{s1}), there exists a $h\in C$ with $h\in W_i$ and $i\in W_h$.  By Lemma \ref{lemcn} (i),   $x_h\in S_{n+1}- S_n$ for some $n\geq 0$. By Lemma \ref{lemcn} (ii) $x_i\in S_n$, hence, $x_i\prec_p x_h$. 

To prove the lemma we show that $x_h \mbox{ controls}_p \ x_i$.
Let  $\{y_j:j\in W_t\}\rightarrow_p y_t=x_h$ be composing, hence, $t\in C$ by 
Lemma \ref{lemcon}. Then (\ref{s2}) with $h\in C$ and $x_h=y_t$ states $\{x_j:j\in W_h\}=\{y_j: j\in W_t\}$. 
 This proves $x_i\in \{y_j: j\in W_t\}$ and we conclude that $x_h \mbox{ controls}_p \ x_i$. 

\hfill
$\Box$

The composing/decomposing theorem is compatible with the knowledge flow theorem if $F_p$ satisfies the single source axiom. We need to show that it is hard to compute an element $v_0\in F_p$.  First, observe that locally collision free sets often satisfy the following condition that is slightly stronger than (\ref{s2}): for each $i\in C$ there exists an injective {\em function} $c_i$ such that $c_i((x_j)_{j\in W_i})= x_i$ for $(x_1,\ldots,x_m)\in S$ and such that the image of $c_i$ has an empty intersection with the images of $c_j$, $j\neq i$; $Im(S)$ is equal to the union of the images of $c_i$, $i\in C$.

Since $F_p\subseteq Im(S)$, $v_0=c_i(x_1,\ldots, x_m)$ for some  $i\in C$. Since $F_p$ is a fixed set, $v_1\in F_p$ for some $v_1=x_j$. Thus, $v_0=q_0(v_1)$ for some function derived from $c_i$. By continuing this argument we obtain a sequence of elements $v_0=q_0(v_1), v_1=q_1(v_2), \ldots$ In practise, the domain of the functions $c_i$, $i\in C$, has a finite size. Thus there exist $j< h$ with $v_j=v_h$, that is, $v_j$ is a fixed point of the equation $x=w(x)$ where $w(x)=q_j(q_{j+1}(\ldots q_{h-1}(x) \ldots ))$ is some function composed of $c_i$, $i\in C$. So, the single source axiom is satisfied if it is hard to compute compositions $q_0(q_1(\ldots q_{j-1}(x)\ldots ))$ with $x=w(x)$.

\bibliographystyle{plain}

\end{document}